\documentstyle[preprint,aps]{revtex}
\begin{document}
\title{{\bf Axial Vector Couplings of the Nucleon in Chiral Quark Model}
{\bf Incorporating $ U(1)_A$ Anomaly Effects}\thanks
{This work is supported in part by funds provided by the China National
Natural Science Foundation.}}
\author{\underline{Xiaoyuan Li}$^1$\thanks{E-mail: Lixy@bepc2.ihep.ac.cn.}
and Yi Liao$^2$\thanks{E-mail: Liaoy@itp.ac.cn.}}
\address{$^1$Institute of Theoretical Physics, Chinese Academy of Sciences,
P. O. Box 2735 Beijing 100080, People's Republic of China\\
$^2$Department of Modern Applied Physics, Tsinghua University, Beijing
100084,\\
People's Republic of China}

\maketitle

\thispagestyle{empty}

\begin{abstract}
Renormalization of the axial vector currents due to Goldstone loops is
studied in a simple extension of Manohar - Georgi chiral quark model
which incorporates $U(1)_A$ anomaly effects. The polarized strage quark
sea in the polarized nucleon results from different renormalization of the
flavor singlet and octet currents and is in reasonable agreement with the
experiment.
\end{abstract}

\vspace*{\fill}
\begin{center}
Submitted to: {\it Physics Letters B}
\end{center}

\begin{flushleft}
\end{flushleft}

\newpage
The chiral quark model of Manohar and Georgi\cite{MG} offers an
explanation of why the nonrelativistic quark (NRQ) model works so well
for hadrons made up of light quarks. The electroweak properties of
constituent quarks such as axial vector couplings and magnetic moments,
however, remain as undetermined coefficients in nonlinear chiral
Lagrangian. Using the above model in the chiral limit, Weinberg\cite{Weinberg}
argued some time ago that to the leading order
in the large $N_{\rm c}$ expansion ($N_{\rm c}$ is the number of
colors), constituent quarks behave like bare Dirac particles, i.e.,
with isospin triplet axial vector coupling $g_A=1$ and anomalous
magnetic moment ${\kappa}=0$. The next to leading order corrections
have also been estimated by sum rules or Feynman diagram calculations
in the nonlinear ${\sigma}$ model or the linear ${\sigma}$ model
using the ${\sigma}$ field as an effective regulator\cite{Peris,wein,DMKV,LL}.

What about the flavor singlet axial vector coupling $g_A^0$ of
the quark? Particular interest in this problem arises when analyzing
the EMC\cite{EMC} and more recent SMC\cite{SMC} and
SLAC\cite{SLAC} measurements of the nucleon spin structure
functions. The original EMC result combined with data on nucleon
and hyperon ${\beta}$ decays suggested that quarks carry only a small
fraction of the nucleon spin and the strange quark sea is substantially
polarized in the direction opposite to the nucleon spin, thus
in conflict with the simple picture of the NRQ model. Recent
experiment results have reduced the polarization effect of the strange
quark sea but increased its statistical significance. As emphasized by
Kaplan and Manohar \cite{KM}, upon understanding this discrepancy,
the key is to remember that the constituent quarks of the quark model
are not the same things as the current quarks of QCD. Thus, when going
from the current to constituent quarks the axial vector currents are
subject to renormalization due to spontaneously broken chiral symmetry
in QCD. Furthermore, this renormalization in the picture of chiral
quark model arises from loops of Goldstone bosons. If the singlet
and octet axial vector currents are renormalized differently, it is
possible to have a polarized strange quark sea in the polarized
nucleon. In this letter we will study the whole three axial vector couplings
$G_A^{3,8,0}$ of the nucleon (i.e., the 3rd and 8th components of the
octet and the singlet) in a simple extension of chiral quark model. We will
see that a polarized strange quark sea so obtained can indeed be in reasonable
agreement with experiment.

Any theoretical approach trying to give a reasonable value for the
flavor singlet coupling $G_A^0$ is faced with the difficulty of
how to incorporate the $U(1)_A$ anomaly effects. In the sector of pure
pseudoscalar nonet it has generally been agreed how to do this in a chiral
and large $N_C$ expansion though
controversies still persist\cite{U(1),Eff}. With the inclusion
of quarks, i.e., in chiral quark model, simply adding the anomaly term
to chiral Lagrangian would double-count the anomaly effects because
the anomaly is still hidden in the quark integral measure when we
quantize quark fields. Fortunately there is a way to avoid the dilemma.
The key point is that one may use a $U(1)_A$-neutral quark field\cite{ES}.
This is possible because the $U(1)_A$ transformation of a quark field is just
a chiral analog of the ordinary phase redefinition. For example, if the quark
field $\psi$ and the flavor singlet pseudoscalar $\eta_0$ transform under
$U(1)_A$ as $\psi \rightarrow \psi '= \exp(i\omega\gamma_5)\psi$ and
$\eta_0 \rightarrow \eta_0 ' = \eta_0 - \omega f_0\sqrt{6}$, one may
define the $U(1)_A$-neutral quark field to be $Q = \exp(i \frac{\displaystyle{
\eta_0\gamma_5}}{\displaystyle{\sqrt{6}f_0}})\psi$. The quark integral measure
in
terms of Q is also neutral and no longer generates $U(1)_A$ anomaly so that
one can now unambiguously include the anomaly terms in the chiral lagrangian
for pseudoscalars,~${\cal L}_{\rm ps}$. For our purpose here the relevant point
is that the $\eta_0$ gets a large mass which is nonzero even in the chiral
limit. Symmetry considerations dictate the form of chiral Lagrangian
at the lowest order in chiral and large $N_C$ expansion
\begin{equation}
\begin{array}{l}
\displaystyle{\cal L}=\bar{Q}(i\rlap/\partial-M)Q
   +\bar{Q}(\rlap/V+g_A \rlap/A \gamma_5+g_A^0 \rlap/A^0 \gamma_5)Q \\
\displaystyle ~~~~-\bar{Q}(\hat{\xi} m \hat{\xi} \tilde{\xi}\tilde{\xi} P_-
   +\hat{\xi}^{\dagger} m \hat{\xi}^{\dagger}
(\tilde{\xi}\tilde{\xi})^{\dagger}
    P_+)Q + {\cal L}_{{\rm ps}},\\
\displaystyle \hat{\xi}=\exp(i\pi^a \lambda^a/2f),
                        ~~~\tilde{\xi}=\exp(i\eta_0 /\sqrt{6}f_0), \\
\displaystyle P_{\pm}=\frac{1}{2}(1\pm \gamma_5),~~~ m = {\rm diag}(0, 0,
m_s)\\
\displaystyle V_{\mu}
   =\frac{i}{2}(\hat{\xi}\partial_{\mu}\hat{\xi}^{\dagger}
                +\hat{\xi}^{\dagger}\partial_{\mu}\hat{\xi}),
{}~~~A_{\mu}
   =\frac{i}{2}(\hat{\xi}\partial_{\mu}\hat{\xi}^{\dagger}
                -\hat{\xi}^{\dagger}\partial_{\mu}\hat{\xi}), \\
\displaystyle
A_{\mu}^0
   =\frac{i}{2}(\tilde{\xi}\partial_{\mu}\tilde{\xi}^{\dagger}
                -\tilde{\xi}^{\dagger}\partial_{\mu}\tilde{\xi})
   =\frac{1}{\sqrt{6} f_0}\partial_{\mu}\eta_0.
\end{array}
\end{equation}

Several remarks are in order. The second term arises from spontaneous
chiral symmetry breaking, giving quark a constituent mass $M$.
The explicit $SU(3)$ breaking is induced by the $m$ term which also gives
rise to symmetry breaking terms in ${\cal L}_{\rm ps}$.
We have ignored the current mass of u and d quarks. The remaining $g_A$
and $g_A^0$
terms account for the rule that in effective field theory we must include
all terms that are allowed by symmetry and of the same order by chiral
power counting. Compared to the Manohar - Georgi chiral Lagrangian
the above Lagrangian contains an additional term proportional to $g_A^0$.
The singlet and octet axial vector couplings of quark,
$g_A^0$, $g_A$ are free parameters in chiral quark model. As the
Weinberg's argument for $g_A=1$ in large $N_c$ limit was later
challenged and there were indications that $g_A=1$ might not be a
necessary result of large $N_c$ QCD\cite{PR,BLS,JZ},
we would like to take $g_A$ as free and calculate its renormalization
effects arising from chiral loops which are believed to be one
of the important contributions at the next order. In the case of $g_A^0$
the situation is more obscure. There is no analogous sum rule to
constrain it as used for $g_A$, so we set it free as well.

We are now ready to compute $G_A^{0,3,8}$, which are defined by the
nucleon matrix elements of the QCD current $\displaystyle
j_{\mu 5}^a=\bar{q}\gamma_{\mu}\gamma_5\frac{\lambda^a}{2}q$ with
$\lambda_a~(a=1-8)$ the Gell-Mann matrices and
$\displaystyle\lambda^0=\sqrt{\frac{2}{3}} {\rm diag}(1~~1~~1)$,
\begin{equation}
\begin{array}{l}
\displaystyle <{\rm N}(p+q)|j_{\mu 5}^0|{\rm N}(p)>=
  \frac{1}{\sqrt{6}}\bar{u}(p+q)[\gamma_{\mu}G_A^0(q^2)
                                 +q_{\mu}H_A^0(q^2)]\gamma_5 u(p), \\
\displaystyle <{\rm N}(p+q)|j_{\mu 5}^3|{\rm N}(p)>=
  \frac{1}{2}\bar{u}(p+q)[\gamma_{\mu}G_A^3(q^2)
                                 +q_{\mu}H_A^3(q^2)]\gamma_5\tau^3 u(p), \\
\displaystyle <{\rm N}(p+q)|j_{\mu 5}^8|{\rm N}(p)>=
  \frac{1}{2\sqrt{3}}\bar{u}(p+q)[\gamma_{\mu}G_A^8(q^2)
                                 +q_{\mu}H_A^8(q^2)]\gamma_5 u(p), \\
\displaystyle G_A^{0,3,8}=G_A^{0,3,8}(0).
\end{array}
\end{equation}
Note that for $G_A^0$ we have to specify its renormalization scale
$\mu \leq \Lambda_{\chi {\rm SB}}$ (chiral symmetry breaking scale).
In the spirit of effective field theory, at scale
$\mu \leq \Lambda_{\chi {\rm SB}}$ we may make the appropriate
substitution $j_{\mu 5}^a \rightarrow J_{\mu 5}^a$, where $J_{\mu 5}^a$
is derived from chiral Lagrangian,
\begin{equation}
\begin{array}{l}
\displaystyle J_{\mu 5}^a
  =\frac{1}{4}\bar{Q}\gamma_{\mu}(\hat{\xi}\lambda^a\hat{\xi}^{\dagger}
                                 -\hat{\xi}^{\dagger}\lambda^a\hat{\xi})Q
  +\frac{g_A}{4}\bar{Q}\gamma_{\mu}\gamma_5
                                 (\hat{\xi}\lambda^a\hat{\xi}^{\dagger}
                                 +\hat{\xi}^{\dagger}\lambda^a\hat{\xi})Q
  +({\rm meson~~terms}), \\
\displaystyle J_{\mu 5}^0=\frac{g_A^0}{\sqrt{6}}
              \bar{Q}\gamma_{\mu}\gamma_5 Q+({\rm meson~~terms}).
\end{array}
\end{equation}
So, effectively we have
\begin{equation}
\displaystyle <{\rm N}|j_{\mu 5}^a|{\rm N}>=
              <{\rm N}|J_{\mu 5}^a|{\rm N}>=
              <{\rm N}|\bar{J}_{\mu 5}^{a'}|{\rm N}>.
\end{equation}
In the above second equality we have included renormalization effects
from chiral loops in the coefficients of $\bar{J}_{\mu 5}^{a'}$ while
the matrix elements themselves are to be evaluated at tree level in
specific quark models. Since $SU(3)$ is explicitly broken
$J_{\mu 5}^0$ and $ J_{\mu 5}^8$ mix under renormalization so that
$a'$ involves components besides $a$. Since isospin is conserved,
$J_{\mu 5}^3$ is renormalized multiplicatively. We use the NRQ model
to evaluate the matrix elements, so $\bar{J}_{\mu 5}^a$ involves
only $U$ and $D$ quarks,
\begin{equation}
\begin{array}{l}
\displaystyle
\bar{J}_{\mu 5}^{0'}=\frac{a_0}{\sqrt{6}}
    (\bar{U}\gamma_{\mu}\gamma_5 U+\bar{D}\gamma_{\mu}\gamma_5 D), \\
\bar{J}_{\mu 5}^{3'}=\frac{a_3}{\sqrt{2}}
    (\bar{U}\gamma_{\mu}\gamma_5 U-\bar{D}\gamma_{\mu}\gamma_5 D), \\
\bar{J}_{\mu 5}^{8'}=\frac{a_8}{2\sqrt{3}}
    (\bar{U}\gamma_{\mu}\gamma_5 U+\bar{D}\gamma_{\mu}\gamma_5 D),
\end{array}
\end{equation}
where $a_{0,3,8}$ are effective axial vector couplings of quark in chiral
quark model and related to observables $G_A^{0,3,8}$ by
\begin{equation}
\displaystyle
G_A^0=a_0,~~G_A^3=\frac{5}{3}a_3,~~G_A^8=a_8.
\end{equation}
Explicit calculation of Feynman diagrams in Fig.~1 shows that $a_{0,3,8}$ have
the following
structure,
\begin{equation}
\begin{array}{l}
\displaystyle
a_0=g_A^0(1-A),\\
a_3=g_A(1-A)-(2B^{\pi}+B^K),\\
a_8=g_A(1-A)-3B^K.\\
\end{array}
\end{equation}
Instead of writing down the lengthy formulae for $A$ and $B^{\pi(K)}$,
we emphasize the following features. $A$ is a sum of terms contributed
by the whole pseudoscalar nonet, while $B^{\pi(K)}$ only receives
contributions from $\pi^{\pm}~(~K^{\pm},~K^0~~{\rm and}~\bar{K}^0)$.
The $\eta'$ contributes to $a_{0,3,8}$ in the same way as the
Goldstone octet does. This is because we have actually treated
$\eta'$ as if it were a Goldstone boson. Although it is guided
by large $N_c$ arguments, numerical analysis will tell us whether
it is a good approximation. The singlet and nonsinglet couplings are
renormalized differently even in the chiral limit and with $g_A =g_A^0$.
This difference arises because Fig.~1(c) and (d) appear only in
the nonsinglet channel. Physically it is responsible for the polarized strange
quark sea in the polarized nucleon,
\begin{equation}
\displaystyle \Delta S =\frac{a_0 -a_8}{3} = \frac{1}{3}(g_A^0-g_A)(1-A)+B^K.
\end{equation}
The splitting between $a_3$ and $a_8$
is due to explicit $SU(3)$ breaking, so in the $SU(3)$ limit we
should have $a_3=a_8$. Indeed, using the 'experiment' value
$a_3=0.75$ and $a_8=0.6$ we estimate that $SU(3)$ breaking effects
are within $30\%$.

To quantify our discussion we regularize as usual ultraviolet
divergences by the cutoff $\Lambda_{\chi {\rm SB}}=4\pi f$,
where $f= 84$ MeV is the decay constant of the Goldstone bosons in the
chiral limit. The relevant input is,
$M=350$ MeV, $m_s=150$ MeV, $m_{\pi}=135$ MeV, $m_K=492$ MeV,
$m_{\eta}=547$ MeV, $m_{\eta'}=958$ MeV, $\theta=- 20^{\circ}~(\eta-\eta'$
mixing angle), $f_{\pi}=f_{\eta}=f_{\eta'}=130/\sqrt{2}$ MeV,
$f_K=160/\sqrt{2}$ MeV. Our discussion does not depend on the
details of the input. Then $G_A^{0,3,8}$ are functions of $g_A$ and
$g_A^0$. As they are sensitive to $g_A$ we choose a number for
$g_A$ so that we may get a not-too-bad number for $G_A^3$ in a
reasonable range of $g_A^0$. The result with $g_A=1.13$ is plotted
in Fig. 2. We see that the global pattern of $G_A^{0,3,8}$
with a positive $g_A^0$ is in reasonable agreement with experiment.
For example, at $g_A = g_A^0 =1.13$, we have $G_A^3 =1.23 $  ,$ G_A^8
=0.45 $    ,
$G_A^0 =0.16  $  and  $ \Delta S =-0.10       $. Considering the simplicity of
our
working Lagrangian this is encouraging. Especially, an important part of the
strange quark polarization can indeed be attributed to different
renormalization
of the singlet and octet axial vector currents by chiral loops.
Although it is possible to
fit all of $G_A^{0,3,8}$ to experiment, this requires a large
negative value for $g_A^0$. (For example, $G_A^0=0.3$,
$G_A^3=1.25$ and $G_A^8=0.6$ using $g_A=1.16$, $g_A^0=-2.14$.)
A negative sign for the ratio $\zeta=g_A^0/g_A$ was also
favored by a recent work based on a quantum-mechanical analysis
in chiral quark model\cite{CL}. But we still think this is not very natural
because it is hard to believe that the next order
correction in a good perturbative expansion would change the value of $G_A^0$
from $-2$ or $-1$ to $0.3~$.

We should mention a weak point in our discussion which seems to deserve
further study. The corrections to $ G_A^{0, 3, 8}$ computed here are dominated
by chiral loops of the Goldstone octet. The contribution of the $\eta '$
is less important even in the singlet channel while intuitively one expects
that the $\eta '$ should couple strongly to the singlet channel. This
occurs because we have actually treat $\eta '$ as a Goldstone boson. We
guess that if we put in somehow the strong coupling between $\eta '$ and
the singlet current we would have a better expansion of the experimental
values of $ G_A^{0, 3, 8}$. This is not totally impossible. An additional
diagram like Fig.~1(c) or (d) would produce some " $B$ " term
which partially cancels the " $A$ " term in $a_0$. Indeed a similar
cancellation
does occur in the octet coupling $a_{3, 8}$. If the guess is really correct we
can start with a smaller $g_A^0$ but end up with larger $G_A^0$ ( hence a more
reliable perturbative treatment in the singlet channel) and $G_A^{3,8}$, thus
in closer agreement with experiment. Inversely, this may imply that some
higher order terms in large $N_C$ expansion of chiral Lagrangian are probably
important. The other point not touched upon in this letter concerns the
running property of the singlet axial vector current in the region
$ \mu < \Lambda_{\chi {\rm SB}} $. According to the analysis in Ref.~\cite{MG}
strong
interactions in this region are much weakened, compared to the naive
extrapolation from the perturbative region of QCD. If we consider our computed
$G_A^0$ to be evaluated at $\mu \approx M$ and mimic the running in the
intermediate region $ \Lambda_{\chi {\rm SB}} > \mu > M $ by simply using a
" scaled-down " QCD ( just as we mimic technicolor by using a " scaled-up "
QCD ) we estimate
$G_A^0(\mu=\Lambda_{\chi {\rm SB}}) \approx 0.98~ G_A^0(\mu= M)$\cite{Liao}.

Finally we argue that $g_A$ appearing in the chiral Lagrangian
depends in some sense on the number of light flavors involved.
Usually we work with two flavors ( u and d ) when we determine from
$g_A$ the coupling $G_A^3$ as measured in the neutron $\beta$ decay.
To determine couplings of other components, $G_A^0$ and $G_A^8$,
we surely have to include the strange quark. But we should obtain
the same $G_A^3$ whether we work with two or three flavors.
To the leading order, the relation between $g_A$ and $G_A^3$
is unchanged, e.g. as in the NRQ model. Its next order corrections
arising from chiral loops are basically determined by quadratic
Casimir operators of the flavor symmetry group in the symmetry limit.
Then one way out is that $g_A$ also depends on the number of flavors
involved. This can be understood in the language of effective field
theory. When we are working with two flavors we have already integrated
out the strange quark, $K$ and $\eta$ fields and inserted their
renormalization effects directly into $g_A$ which is a parameter
in chiral Lagrangian with two flavors. This explains why we required
a larger value of $g_A$ than usual to fit $G_A^3$.

  We thank Y. P. Kuang for helpful discussions.

\null
\newpage
\centerline{\large\bf Figure Caption }
Fig.~1  Feynman diagrams contributing to renormalization of axial vector
currents. Thick lines and thin lines represent quarks and pseudoscalars
respectively. Solid circles represent insertion of current.

Fig.~2 The computed couplings $3/5 G_A^3,~G_A^8$ and $G_A^0$ ( upper, middle
and
lower curves) of the nucleon are shown as functions of $g_A^0$ at
$g_A=1.13$.

\null
\newpage
\input FEYNMAN
\textheight 800pt \textwidth 450pt
\begin{picture}(10000,18000)

\THINLINES\put(14000,-10000){\oval(4000,4000)[t]}
\THICKLINES\drawline\fermion[\E\REG](14000,-10000)[4000]
\THICKLINES\drawline\fermion[\W\REG](14000,-10000)[4000]
\put(17000,-10000){\circle*{500}}
\put(13200,-12000){(a)}

\THINLINES\put(30000,-10000){\oval(4000,4000)[t]}
\THICKLINES\drawline\fermion[\E\REG](30000,-10000)[4000]
\THICKLINES\drawline\fermion[\W\REG](30000,-10000)[4000]
\put(30000,-10000){\circle*{500}}
\put(29200,-12000){(b)}

\THINLINES\put(14000,-20000){\oval(4000,4000)[t]}
\THICKLINES\drawline\fermion[\E\REG](14000,-20000)[4000]
\THICKLINES\drawline\fermion[\W\REG](14000,-20000)[4000]
\put(16000,-20000){\circle*{500}}
\put(13200,-22000){(c)}

\THINLINES\put(30000,-18000){\circle{4000}}
\THICKLINES\drawline\fermion[\E\REG](30000,-20000)[4000]
\THICKLINES\drawline\fermion[\W\REG](30000,-20000)[4000]
\put(30000,-20000){\circle*{500}}
\put(29200,-22000){(d)}

\end{picture}

\vspace{7.5cm}
\begin{center}
Fig.~1~~~~~Xiaoyuan Li, Phys. Lett. B\\
\end{center}
\end{document}